# Can Microsoft Academic help to assess the citation impact of academic books?[1]

Kayvan Kousha and Mike Thelwall

Statistical Cybermetrics Research Group, School of Mathematics and Computer Science, University of

Wolverhampton, Wulfruna Street, Wolverhampton WV1 1LY, UK.

Despite recent evidence that Microsoft Academic is an extensive source of citation counts for journal articles, it is not known if the same is true for academic books. This paper fills this gap by comparing citations to 16,463 books from 2013-2016 in the Book Citation Index (BKCI) against automatically extracted citations from Microsoft Academic and Google Books in 17 fields. About 60% of the BKCI books had records in Microsoft Academic, varying by year and field. Citation counts from Microsoft Academic were 1.5 to 3.6 times higher than from BKCI in nine subject areas across all years for books indexed by both. Microsoft Academic found more citations than BKCI because it indexes more scholarly publications and combines citations to different editions and chapters. In contrast, BKCI only found more citations than Microsoft Academic for books in three fields from 2013-2014. Microsoft Academic also found more citations than Google Books in six fields for all years. Thus, Microsoft Academic may be a useful source for the impact assessment of books when comprehensive coverage is not essential.

## 1. Introduction

Edited books and monographs are important academic outputs in the arts and humanities and some social sciences (Nederhof, 2006; Huang & Chang, 2008; Hammarfelt, 2016). For instance, about a third of research publications from Australian universities in the social sciences and humanities two decades ago were books or book chapters (Bourke & Bulter, 1996) and the proportion of book submissions to the 2008 UK Research Assessment Exercise (RAE) across 38 social sciences and arts and humanities subject areas was 31% (Kousha, Thelwall, & Rezaie, 2011). Prior to the creation of the first major book citation index, citation impact monitoring for books was more challenging than for journal articles (Garfield, 1996). This was because, in many humanities and some social science fields, books attract more citations from other books than from journal articles. Bibliometric indicators based on journal-based citation indexes therefore do not fully reflect the intellectual impact of books (Cronin, Snyder, & Atkins, 1997; Hicks, 1999; Archambault et al., 2006). In political science, for example, one study found that books received almost three times more citations from other books than from Web of Science (WoS) journal articles (Samuels, 2013) and another found that Google Books citations to academic books were more common than Scopus citations in the humanities (Kousha, Thelwall, & Rezaie, 2011).

Lacking a book citation index, some early investigations manually checked references in scholarly documents (e.g., Cullars, 1998; Knievel & Kellsey, 2005; Krampen, Becker, Wahner, & Montada, 2007; Creaser, Oppenheim, & Summers, 2011), used the cited reference search facility in WoS to count citations to books (e.g., Butler & Visser, 2006; Bar-Ilan, 2010; Hammarfelt, 2011; Chi, 2014) or used non-citation indicators (e.g., library holdings: White, Boell, Yu, Davis, Wilson, & Cole, 2009) to assess the broader impacts of books (for reviews see: Kousha & Thelwall, 2015; Hammarfelt, 2016). Google Scholar or Google

---

[1] Kousha, K & Thelwall, M. (in press, 2018). Can Microsoft Academic help to assess the citation impact of academic books. Journal of Informetrics.

Books citation searching can also find citations from books or other publications that are absent from WoS and Scopus (Kousha & Thelwall, 2009; Kousha, Thelwall & Rezaie, 2011; Abdullah & Thelwall, 2014). These methods are problematic to apply in practice for large-scale systematic citation analyses of book chapters or monographs because (except perhaps for Google Books) they are not comprehensive enough (Giménez-Toledo et al., 2016) and Google Books citations only include citations from books (Kousha & Thelwall, 2009).

Thomson Reuters (now Clarivate Analytics) introduced the Book Citation Index (BKCI) in 2011, adding citations from books to the WoS interface for an additional charge. This is not yet a perfect solution because BKCI citation counts can be underestimates for books published in multiple editions and for edited volumes (Leydesdorff & Felt, 2012; Gorraiz, Purnell, & Glänzel, 2013; Glänzel, Thijs, & Chi, 2016) and BKCI indexes relatively few books, and very few non-English works (Gorraiz, Purnell, & Glänzel, 2013; Torres-Salinas et al., 2014).

Microsoft Academic is a relaunched free academic citation index that has indexed over 175 million scholarly publications, including from 48,000 journals and 4,000 conferences (https://academic.microsoft.com/ as of June 2018). It captures more citations to journal articles than WoS and Scopus (Harzing & Alakangas, 2017a; Hug & Brändle, 2017; Hug, Ochsner, & Brändle, 2017; Thelwall, 2017; Kousha, Thelwall, & Abdoli 2018). Microsoft Academic also indexes books (Hug & Brändle, 2017) and may also extract citations from them, especially if they are open access. It supports automatic searches, allowing accurate large-scale citation analyses (Hug, Ochsner, & Brändle, 2017; Thelwall, 2018b). Thus, Microsoft Academic seems likely to be useful for the research impact assessment of academic books. To investigate this, the current article compares Microsoft Academic citations with BKCI and Google Books citations to 16,463 BKCI books in 17 fields.

## 2. Databases for book citation counting

### 2.1. The Book Citation Index

By early 2018, BKCI included over 60,000 books from 2005, covering Social & Behavioral Sciences and the Arts & Humanities (60%) and Natural Sciences (40%)[2]. Most indexed books are in English (97%) and published in the UK or the USA (75%) (Torres-Salinas et al., 2014), which is problematic for counting citations to non-English books. For instance, only 4% of German political scientists' books had been indexed by BKCI (Chi, 2014). BKCI claims that it combines citations from core WoS publications (mostly journal articles and conference papers) with citations from the BKCI-indexed books. However, most citations to BKCI books still come from journal articles (92% in sciences and 80% in social sciences and humanities) rather than books (5% and 16% respectively) (Kousha & Thelwall, 2014). Thus, the current version of BKCI does not seem to index enough academic books to make a difference for book impact assessments.

### 2.2. Scopus books

In 2013 Elsevier initiated the Scopus Book Titles Expansion Program[3] to add scholarly books to its main database of journal articles and conference papers. The Scopus advanced search command "DOCTYPE(bk)" can be used to retrieve a list of academic books by different individuals, institutions, or countries. Although Scopus indexes twice as many as academic books (over 150,000[4]) as BKCI, it lacks an effective classification scheme, which is a serious limitation for citation impact assessment. For instance, Scopus only uses one broad category for "Arts and Humanities" and "Social Sciences" and the Journal Classification Codes in Scopus (ASJC) that are designed for retrieving journal articles in narrow fields seem to be rarely used for books. For example, the query "DOCTYPE(BK) AND SUBJMAIN(1203)" for Language and Linguistics books (ASJC code 1203), only retuned five matches from the entire Scopus database, although BKCI had indexed several thousand books in this category (Linguistics; Language & Linguistics). Moreover, Scopus may also be unable to match many citations with its indexed books. For instance, the 2013 book "*Spoken Corpus Linguistics: From Monomodal to Multimodal*" by Svenja Adolphs that was indexed by both BKCI and Scopus had 23 citations in BKCI but no Scopus citations, whereas a Scopus cited reference search found 26 of its citations.

---

[2] http://wokinfo.com/products_tools/multidisciplinary/bookcitationindex/
[3] https://www.elsevier.com/about/press-releases/science-and-technology/elsevier-announces-its-scopus-book-titles-expansion-program

### 2.3. Google Scholar

Google Scholar does not claim to be a book citation index, but it links citations from its databases to books indexed by Google Books, and seems to incorporate citations from Google Books. Google Scholar covers more scholarly-related publications than WoS or Scopus (Khabsa & Giles, 2014; Halevi, Moed, & Bar-Ilan, 2017) and hence could be valuable for book impact assessment, especially from non-Western publishers (Abdullah & Thelwall, 2014). For instance, the 2016 book "*Gaslight Melodrama: From Victorian London to 1940s Hollywood*" by Guy Barefoot has been cited 31 times in Google Scholar. Of these, 11 citations were from other books indexed by Google Books (books.google.com). However, 21 of the citations are to the 2001 edition of the book, indicating that Google Scholar includes citations to other editions. Combining citations to different book editions is a controversial bibliometric issue (Gorraiz, Purnell, & Glänzel, 2013). The main practical limitation of Google Scholar is that it has no API and therefore automatic searches for individual publications are not possible for large-scale book assessments.

### 2.4. Google Books

Google Books contains a substantial number of fully searchable books. Although it does not index citations from books, such citations can be found with an appropriate set of queries and filters (Kousha & Thelwall, 2009). Google Books includes more citations to academic books than Scopus (Kousha, Thelwall, & Rezaie, 2011) and BKCI (Kousha & Thelwall, 2014) in many book-based fields. For example, Google Books citations to academic books submitted to the 2008 UK RAE were three times more numerous than Scopus citations in Law, History, and Communication, Cultural and Media Studies. In contrast to Google Scholar, it is possible to automatically and accurately count citations to books via the Google Books API (Kousha & Thelwall, 2014). This makes Google Books a practical tool for monitoring the citation impact of books. Nevertheless, Google Books does not include citations from journal articles and conference papers to academic books, which is an important disadvantage in many scientific and medical fields.

### 2.5. Microsoft Academic

Microsoft Academic found more citations to a range of document types than Scopus and WoS in multiple fields, including Engineering, Social Sciences, and the Humanities in one study (Harzing & Alakangas, 2017a) and slightly more (6%) citations overall than Scopus to journal articles (Thelwall, 2017). Unlike WoS and Scopus, Microsoft Academic also indexes some preprint archives (Thelwall, 2018a), accounting for its extra citations. The citation advantage of Microsoft Academic over Scopus is also partly due to capturing citations to in-press articles through faster citation indexing (Kousha, Thelwall, & Abdoli, 2018. One small study reported that Microsoft Academic covered 1.5-2.2 times more book chapters, 2.4-4.6 times more monographs and 4-6 times more edited volumes for publications deposited in the University of Zurich Open Archive and Repository (Hug & Brändle, 2017). Nevertheless, there has been no systematic study of Microsoft Academic's coverage of scholarly books or its citation counts for typical books. A practical advantage of Microsoft Academic over Google Scholar is automatic data collection for citation analysis (Harzing & Alakangas, 2017b).

---

[4] https://www.elsevier.com/solutions/scopus/how-scopus-works/content

## 3. Research questions

The main aim of this study is to systematically assess the value of Microsoft Academic automatic searches for the citation impact assessment of academic books. There is no comprehensive list of published academic books and so BKCI is used to give a large, but admittedly biased, sample classified by field. The following research questions drive the investigation.

1. What proportion of academic books from BKCI can be found by Microsoft Academic automatic searches, and does this vary by field and publication year?
2. Does Microsoft Academic find more citations than BKCI and Google Books to the BKCI academic books in its index?
3. Do Microsoft Academic citations to academic books correlate with their BKCI and Google Books citations?

## 4. Methods

In this study only BKCI, Microsoft Academic and Google Books were used to identify citations to academic books. Google Scholar does not support automatic searches and hence it was not practical to use it for a large-scale citation analyses of books. Although Scopus had indexed twice as many English books from 2013-2016 (about 47,000), its subject classification was much less fine grained than that of WoS and so the WoS system was used. Scopus records were not used because further matching of the BKCI books with Scopus books could reduce total number of books in each of the years and fields, widening the statistical confidence intervals produced.

### 4.1. The BKCI data set

BKCI was used for the book sample. Metadata was extracted from 27,989 books published during 2013-2016 in the Book Citation Index-Science (BKCI-S) and Book Citation Index-Social Sciences & Humanities (BKCI-SSH) after data cleaning (see below). These years were selected to assess the influence of time on citation counts in different fields and citation databases. Only English language books were retained to have a more uniform data set for analysis, given that the BKCI coverage of non-English books is already known to be very low. Non-book materials (e.g., articles, biographical items, and reviews) were excluded by selecting "Books" in the "Document Types" option. Because BKCI classifies books into 252 WoS subject categories, which is too fine-grained for statistical analyses (too few books per year and category, 28 on average), the OECD classification scheme was used to combine the books into 17 broad subjects[5]. For instance, Arts, Architecture; Music, Theater, Film, Radio and Television in BKCI were combined to form the broad subject "Art". Similarly, all related medical (e.g., Oncology; Hematology; Surgery), engineering (e.g., Electrical & Electronic Engineering; Materials Science; Civil Engineering), biological (e.g., Cell Biology; Microbiology; Biochemistry & Molecular Biology) and environmental (Environmental Sciences; Geology; Oceanography) books were combined into broad categories.

---

[5] http://www.oecd.org/science/inno/38235147.pdf

## 4.2. Data cleaning

Books with one, two or three words in their titles were excluded to avoid retrieving false matches in Microsoft Academic or Google Books searches (e.g., "Neurovirology", "Systems Biology" or "Advances in Genetics"). Moreover, books without author names or with "Anonymous" authors were removed because authors were necessary for searching and/or filtering the results in Microsoft Academic and Google Books. Book series with volume information such as "Advances in Virus Research, Vol 89" were also removed as far as possible because bibliometric analyses of volume series could be problematic across different citation databases (Leydesdorff & Felt, 2012; Gorraiz, Purnell, & Glänzel, 2013; Glänzel, Thijs, & Chi, 2016). Edition information at the end of book titles was deleted to generate more effective Microsoft Academic searches. For instance, the BKCI title search "*Dental Implant Complications: Etiology, Prevention, and Treatment, 2nd Edition*" gave no results in Microsoft Academic but "Dental Implant Complications: Etiology, Prevention, and Treatment" returned 21 citations. To count citations to the same book editions, their BKCI publication years were matched against Microsoft Academic (see below).

## 4.3. Microsoft Academic citation search

The Microsoft Academic API allows 10,000 free queries per month[6]. Adapting practice for journal articles (Thelwall, 2017; Thelwall, 2018b; Kousha, Thelwall, & Abdoli, 2018), the *Webometric Analyst* free software (http://lexiurl.wlv.ac.uk) was used to run automatic Microsoft Academic citation searches for the BKCI books sample. The Microsoft Academic queries were generated for all 27,989 books by querying their titles, as shown in the example below. Webometric Analyst changes some characters in book titles and uses lowercase letters based on Microsoft Academic's indexing strategy. Only books titles were queried, with subsequent author and publication year filtering to remove false matches. This strategy gives maximal recall and precision, as previously tested for journal articles. Author names were not added to the query because they may remove correct matches (see Thelwall, 2018b).

*Ti='language and time a cognitive linguistics approach'*

The author filtering strategy was important because some books had a shared title (e.g., "*Encyclopedia of the solar system*") but different authors (e.g., "Tilman Spohn" or "John Goodier"). Moreover, Microsoft Academic automatic searches sometimes retrieved book reviews published in journals with the same titles as the books reviewed but with different authors. Book reviews are important scholarly outputs and can potentially be cited (Zuccala & van Leeuwen, 2011). For instance, Microsoft Academic found three records for the book "*What Is Islam?: The Importance of Being Islamic*", including one for the book by Shahab Ahmed and two book reviews in different journals. Hence, extra filtering was needed to remove book reviews from the search results. For this, in addition to author matching, any results with a journal name in the Microsoft Academic "Journal Full Name" field were removed.

Another important filtering strategy was matching the BKCI book publication years with Microsoft Academic search output to match the same book editions (if any) in the two databases. This is important because BKCI reports citations to different book editions separately (Glänzel, Thijs, & Chi, 2016) but Microsoft Academic sometimes merges different book editions. For instance, a Microsoft Academic automatic search for the book "*Leadership in organizations: current issues and key trends*" found 75 citations, but these were for an earlier edition of the book published in 2004 rather than the 2016 edition in BKCI. These early edition cases were excluded to enable fairer comparisons between BKCI and Microsoft Academic citations to books with multiple editions.

---

[6] https://azure.microsoft.com/en-gb/pricing/details/cognitive-services/academic-knowledge-api/

### 4.4. Google Book citation search

Google Books API searches were performed via Webometric Analyst to identify citation to books and remove false matches from the results. Google Books queries were constructed for all books in the data set based on the last name of the first author or editor, as in the BKCI data, a phrase search for the book title, and the publication year (see an example below). False matches filtered out using a set of previously constructed heuristics (Kousha & Thelwall, 2014).

*Lipscomb "Exploring Evidence-Based Practice Debates and Challenges in Nursing" 2016*

## 5. Results

### 5.1. RQ1: Microsoft Academic coverage of BKCI Books

Most (59%; 16,463/27,989) of the BKCI books were found by Microsoft Academic using the above searching and filtering strategy. Microsoft Academic found a higher percentage of BCKI books in science fields, such as *Computer Science* (79%), *Chemical Sciences* (77%) and *Physics and Mathematics* (72%), than in the arts and humanities, such as *Art* (48%), *History and Archaeology* (50%) and *Philosophy, Ethics and Religion* (55%). In social sciences, Microsoft Academic's coverage ranged from 50% in *Educational Sciences* and *Other Social Sciences* to 60% in Sociology and Political Science. Microsoft Academic found more books published in 2013 (64%) and 2014 (63%) than books published more recently in 2015 (58.5%) and 2016 (49%), although the difference is lower in most science fields (Figure 1).

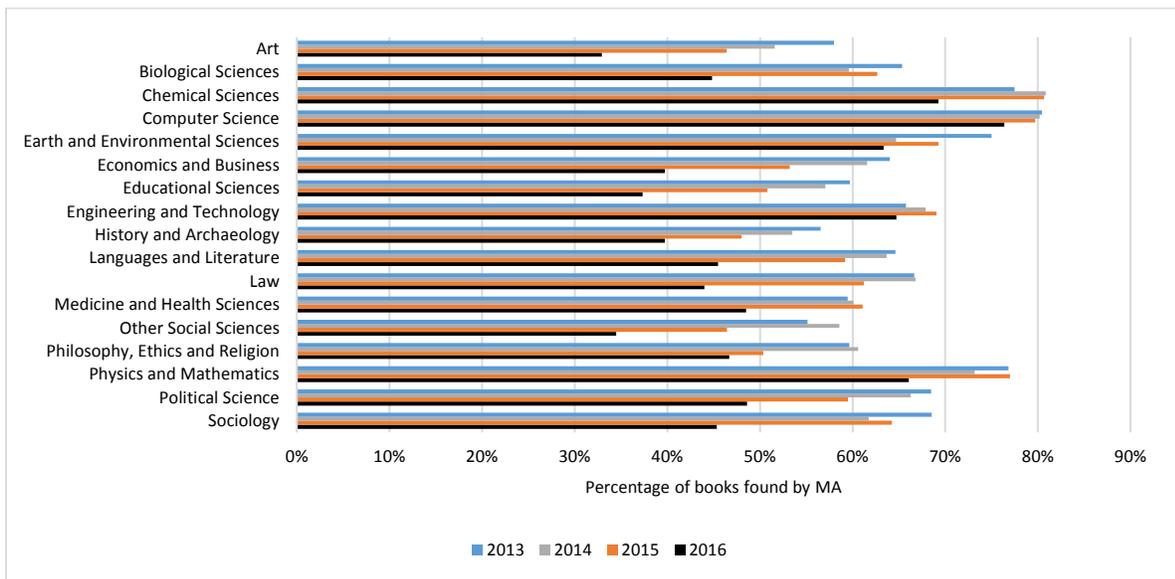

Figure 1. Percentage of BKCI books found by Microsoft Academic by publication year and field.

### 5.2. RQ2: Microsoft Academic citations vs. BKCI and Google Books citations

Figures 2-5 show the average (geometric mean) number of Microsoft Academic, BKCI and Google Books citations to the 16,463 books in 17 fields published during 2013-2016 that were indexed both BKCI and Microsoft Academic. Geometric means were used to compare the average number of Microsoft Academic, BKCI and Google Books citations to books because the arithmetic mean is not a proper central tendency indicator for highly skewed data and the median is also not suitable to differentiate between citation counts with many zeros in the data sets (Thelwall, 2016b). For simplicity, the difference between two averages was judged to be statistically significant if the 95% confidence intervals did not overlap, although a small overlap is also consistent with statistical significance. The issue of spurious positive results due to multiple tests was ignored since the results for individual fields are not critical.

### 5.2.1. Average Microsoft Academic and BKCI citations

Microsoft Academic found statistically significantly more citations than did BKCI in 9 out of 17 fields in all four years (*Chemical Sciences, Medicine and Health Sciences, Physics and Mathematics* from science; *Economics and Business, Political Science, and Other Social Sciences* from the social sciences and *Languages and Literature, Law, Philosophy, Ethics and Religion* from the humanities). The geometric mean number of Microsoft Academic citations was 1.5-3.6 times higher than for BCKI citations in these fields. The geometric mean for Microsoft Academic was also more than for BKCI in *Art* for books published in 2015 and 2016 as well as for *Earth and Environmental Sciences* for all years except in 2014. Surprisingly, Microsoft Academic identified up to three times more citations than BKCI for 2016 books in several subjects (*Economics and Business* (3.6); *Political Science* (3.5); *Physics and Mathematics* (3.4); *Other Social Sciences* (3.3); *Economics* (3.6), Law (3.2); *Chemical Sciences* (3.2); *Medicine and Health Sciences* (3.0); *Philosophy, Ethics and Religion* (3.0); *Art* (3.0). In contrast, the average citation counts from BKCI were not statistically significantly higher than Microsoft Academic in any field or year except for Sociology and History and Archaeology for 2013 and 2014 and Computer Science for 2013.

### 5.2.2. Average Microsoft Academic and Google Books citations

Average Microsoft Academic citation counts were higher than Google Books in 6 out of 17 fields in all four years. This difference is statistically significant at the 95% level in *Chemical Sciences, Physics and Mathematics, Political Science, Other Social Sciences, Languages and Literature, Philosophy, Ethics and Religion* for all years. In contrast, the Google Books geometric means were higher than Microsoft Academic in four subject areas during 2013-2016: *Computer Science; Sociology; History and Archaeology; Educational Sciences* (except for 2013). For 2015-2016, Google Books has also a clear citation advantage over both Microsoft Academic and BKCI in seven fields: *Art; Biological Sciences; Computer Science; Educational Sciences; Engineering and Technology; History and Archaeology; Sociology*. Hence, Microsoft Academic seems to have a lower citation advantage over Google Books compared with BKCI (see above).

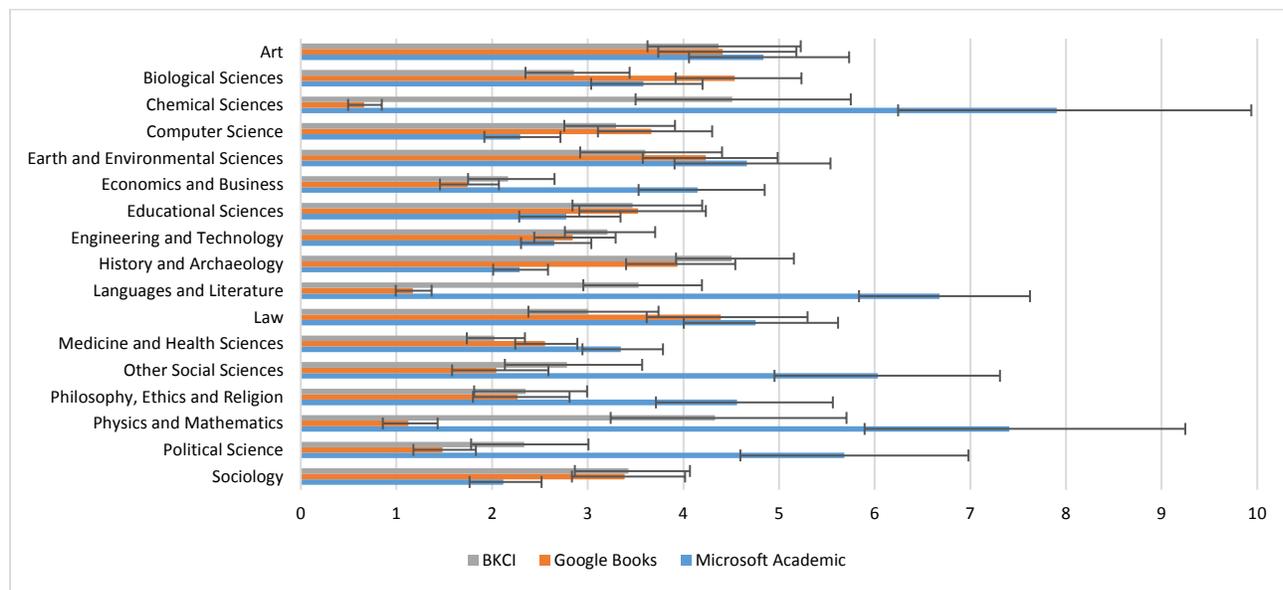

Fig. 2. Geometric mean number of Microsoft Academic, Google Books and BKCI citations and 95% confidence intervals for 2013 academic books across 17 fields.

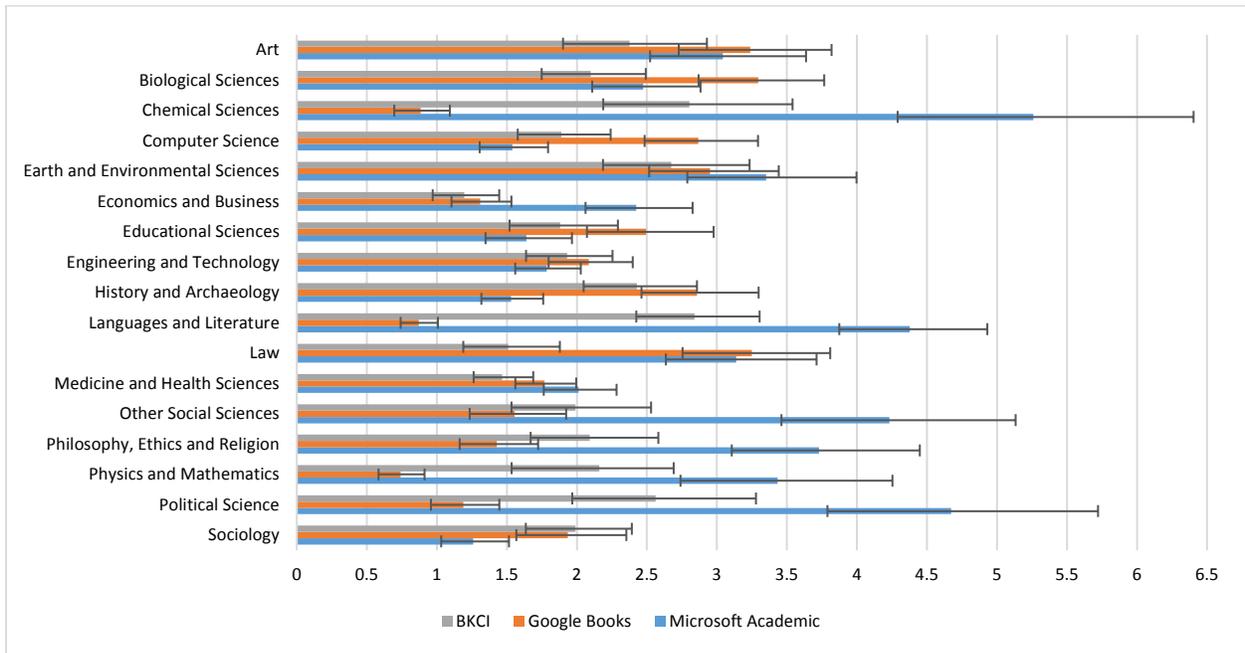

Fig. 3. Geometric mean number of Microsoft Academic, Google Books and BKCI citations and 95% confidence intervals for 2014 academic books across 17 fields.

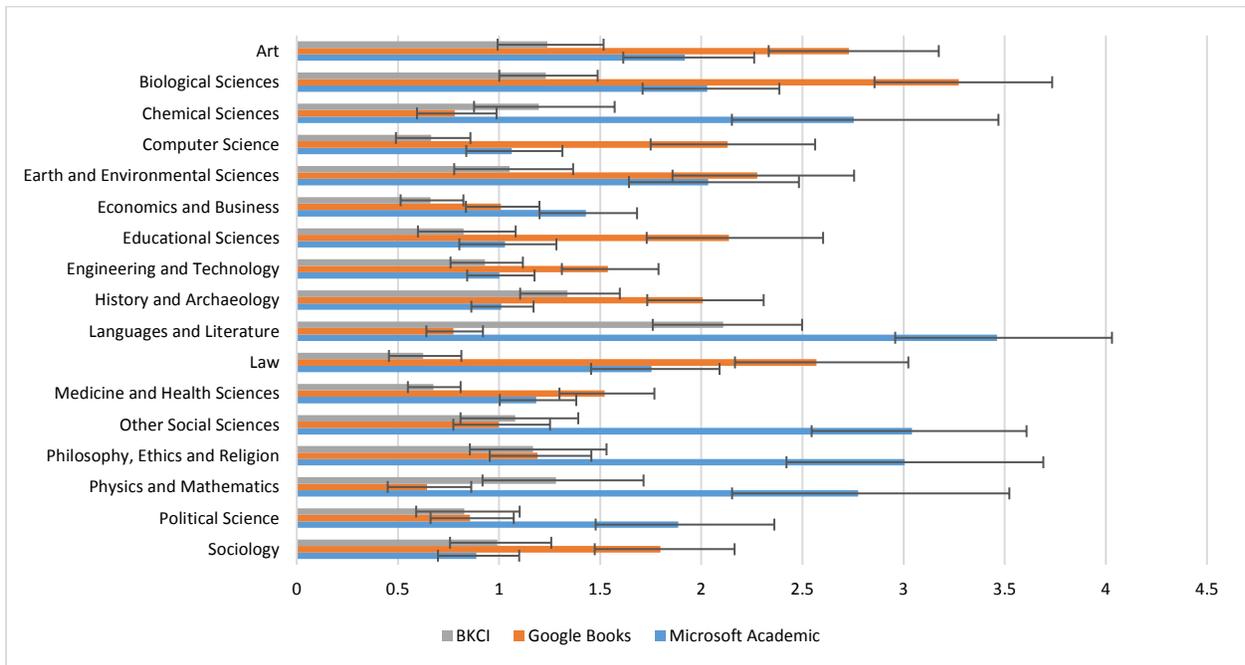

Fig. 4. Geometric mean number of Microsoft Academic, Google Books and BKCI citations and 95% confidence intervals for 2015 academic books across 17 fields.

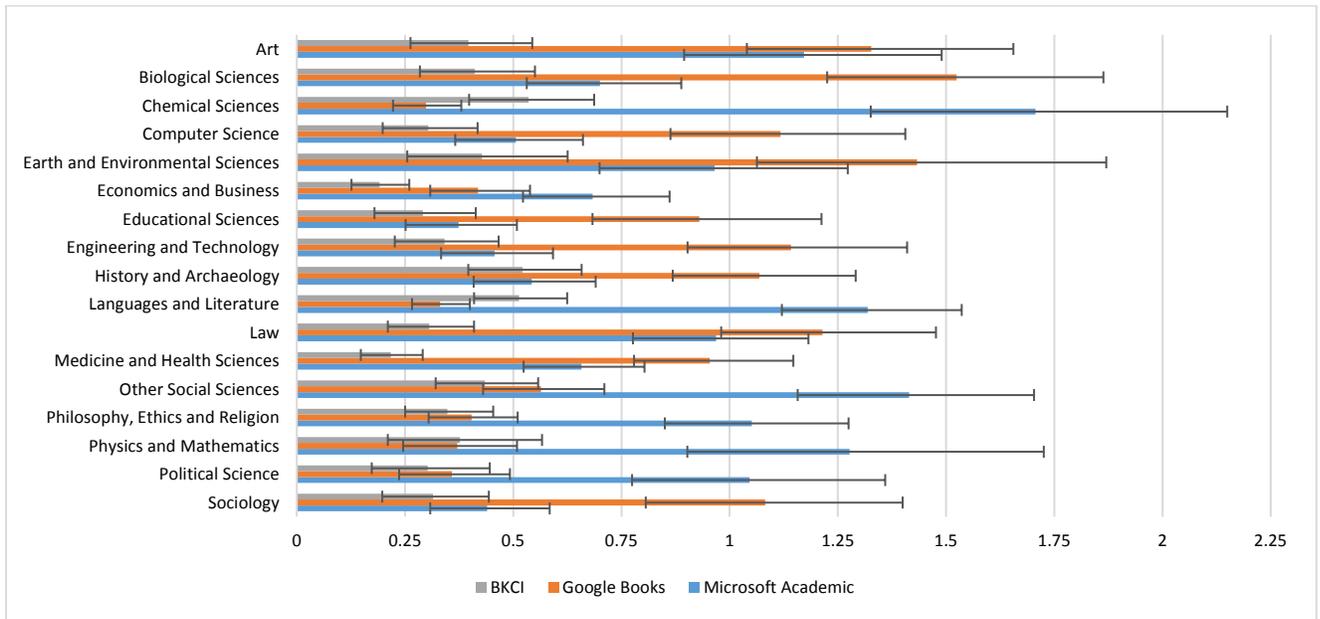

Fig. 5. Geometric mean number of Microsoft Academic, Google Books and BKCI citations and 95% confidence intervals for 2016 academic books across 17 fields.

## 5.3. RQ3: Correlations between Microsoft Academic and BKCI or Google Books citations

There are significant positive Spearman correlations between the Microsoft Academic and BKCI citation counts in all fields and years at the p=0.01 level except for *Philosophy, Ethics and Religion*, and *Other Social Sciences* in 2016 (Figures 6-9). The correlations for science fields such as *Physics and Mathematics* (ranging from .482 to .726 for 2016 and 2013 respectively), *Chemical Sciences* (.447 to .661), *Engineering and Technology* (.439 to .643), *Computer Science* (.376 to 635) and *Biological Sciences* (.440 to .585) are higher than for the other fields, perhaps because in these fields journal and conference citations are more common and both databases have similar coverage of core science and medicine journals (Harzing & Alakangas, 2017a). As mentioned above, in the sciences most BKCI citations are from articles indexed by WoS databases (92%) rather than books indexed by BKCI (about 5%) (Kousha, & Thelwall, 2015).

The correlations between Microsoft Academic and Google Books citations are slightly higher than the correlations between Microsoft Academic and BKCI citations for many social sciences, the arts, and humanities for 2015 and 2016, including *Arts*, *Philosophy, Ethics and Religion*, *Sociology* and *Political Science* (Figures 8-9). This suggests that Microsoft Academic and Google Books broadly reflect similar types of intellectual impact in these fields, perhaps because Microsoft Academic has wider coverage of books (Hug & Brändle, 2017) and preprint archives, such as the *Social Science Research Network (SSRN)* (Thelwall, 2018). The lowest Spearman correlations mostly occurred between BKCI and Google Books citation counts, suggesting that they reflect different types of impacts (mainly article-based impact for BKCI and book-based impact for Google Books).

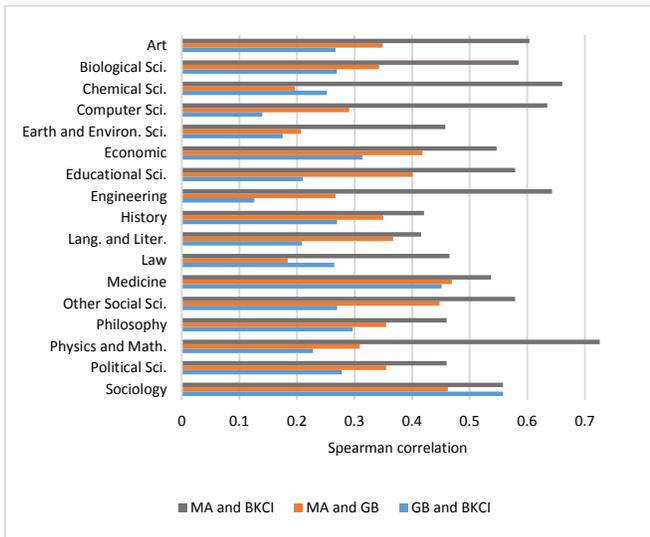
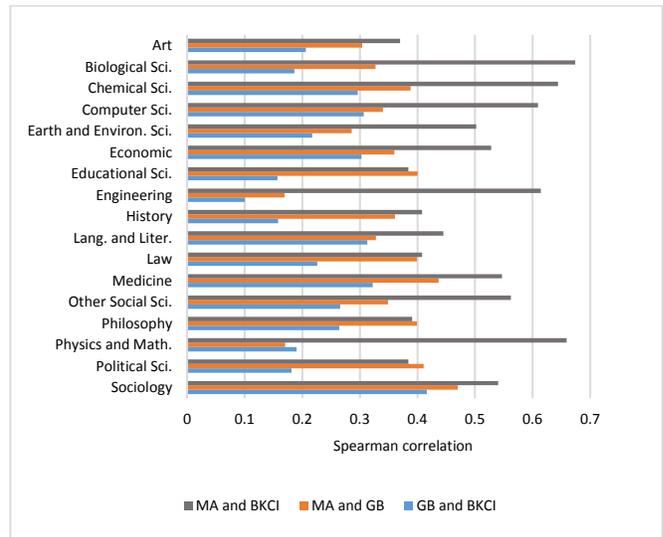

**Figure 6.** Spearman correlations between Microsoft Academic, BKCI and Google Books citations for 2013 academic books across 17 fields.

Figure 7. Spearman correlations between Microsoft Academic, BKCI and Google Books citations for 2014 academic books across 17 fields.

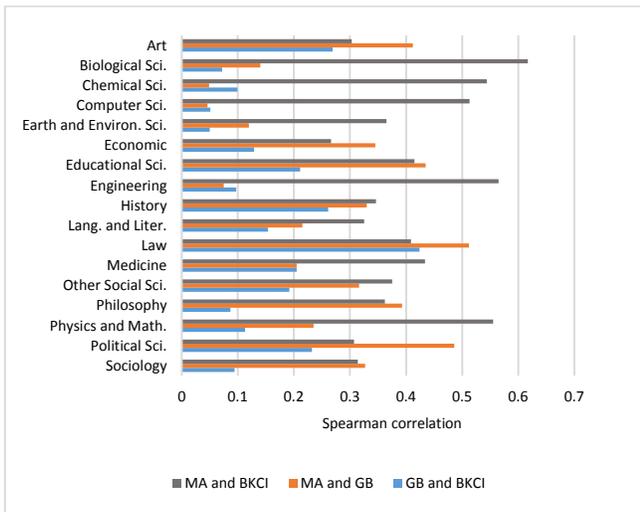
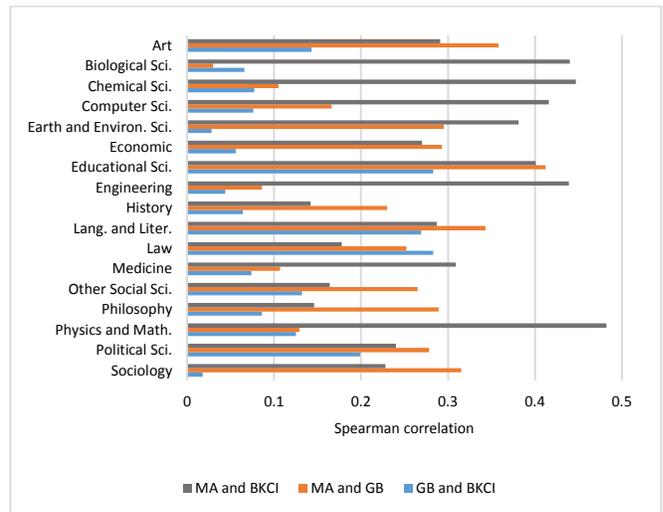

Figure 8. Spearman correlations between Microsoft Academic, BKCI and Google Books citations for 2015 academic books across 17 fields.

Figure 9. Spearman correlations between Microsoft Academic, BKCI and Google Books citations for 2016 academic books across 17 fields.

There are higher correlations between the Microsoft Academic and BKCI citation counts for older books published in 2013 across 13 fields (Figure 10). The higher association for older books may be due to the increasing number of citations over time (e.g., Thelwall, 2016a) and this is clearest for Arts books.

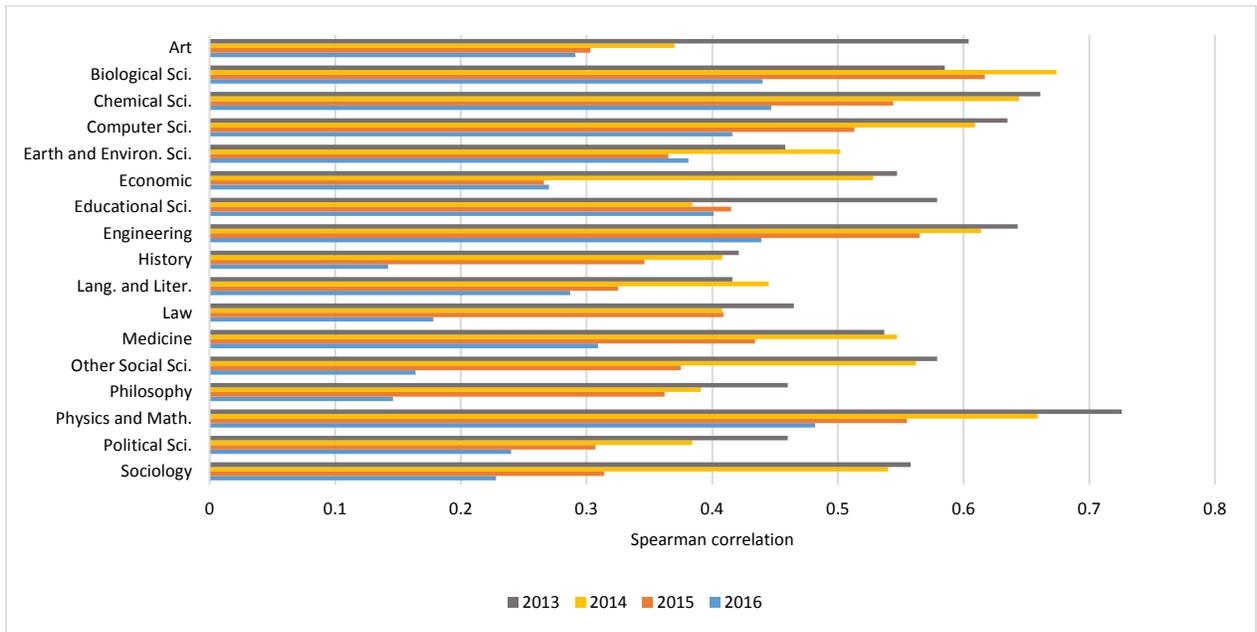

Figure 10. Spearman correlations between Microsoft Academic and BKCI citations to academic books for each individual field, by publication year.

## 6. Discussion

### 6.1. Citations found by Microsoft Academic but not BKCI

A further analysis was conducted to check the overlap between Microsoft Academic citations and BKCI citations to the 15 books in each of the 17 fields (n=255) with the most Microsoft Academic citations but no BKCI citations to find reasons why Microsoft Academic could find more citations. For this, 7,570 sources of citations found by Microsoft Academic to 255 books were extracted and searched in the WoS Core Collection.

#### 6.1.1. Microsoft Academic unique citations

Most 58% (n=4,372) of the Microsoft Academic-indexed citations to 225 books were not found in WoS. For instance, Microsoft Academic identified 21 citations to the 2014 book "*Shakespeare on the university stage*" by Andrew James Hartley but BKCI found none. Manual checks confirmed that all citations found by Microsoft Academic were not indexed in WoS or BKCI. These citations were from journals (e.g., *Shakespeare Bulletin*), books and book chapters. The latter two show that Microsoft Academic can extract citations *from* books, although it is not clear how many citations from books it has indexed. A Scopus cited reference search for the book (REF ("shakespeare on the university stage") found only three citations and so most Microsoft Academic citations were from publications also not indexed by Scopus. Microsoft Academic found more unique citations to books outside WoS in the social sciences, arts and humanities, including *History and Archaeology* (76%), *Art* (73%), *Languages and Literature* (71%) and *Law and Sociology* (both 70%) than for most science fields, such as *Physics and Mathematics* (43%), *Chemical Sciences* (43%) and *Engineering and Technology* (45%) (Figure 11). Thus, Microsoft Academic's main book citation advantage over the current version of BKCI is its greater coverage of citing publications, especially in some arts and humanities fields where BKCI or WoS has relatively weak coverage.

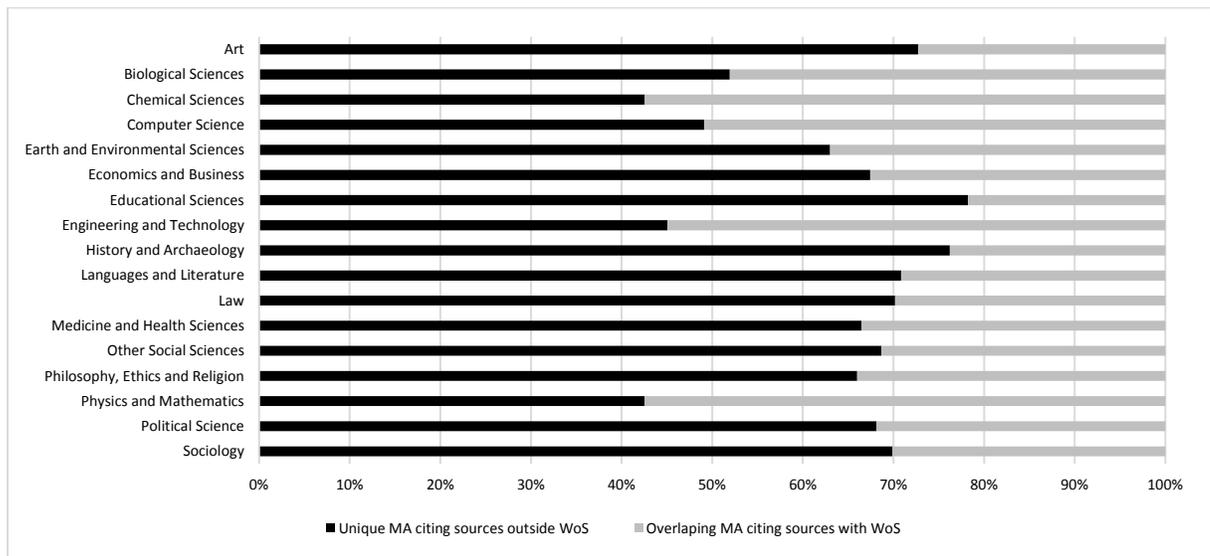

Figure 11. The percentage of unique and duplicate sources of Microsoft Academic citations compared to WoS based on the 15 books with the most Microsoft Academic citations but no BKCI citations for each field.

### 6.1.2. BKCI missed citations

Less than half (42%; n=3,198) of the Microsoft Academic citations to 225 books without BKCI citations were in WoS (Figure 11). For instance, the 2015 book "*Beyond Religious Freedom: The New Global Politics of Religion*" by Elizabeth Shakman Hurd had 28 Microsoft Academic citations but no BKCI citations. This book was cited 57 times in the WoS Core Collection (Example 1 in Table 1) but these citations had not been integrated into BKCI. Extra manual checks showed that many highly cited books in Microsoft Academic without BKCI citations were cited in WoS (highlighted titles in examples 2 and 3 in Table 1). In some of these cases, "*Title: [not available]*" or a series title were listed as the main title in the WoS cited sources. This confirms previous arguments that citations to different book series are not always included in BKCI, creating problems in book impact assessment (Leydesdorff & Felt, 2012; Gorraiz, Purnell, & Glänzel, 2013; Glänzel, Thijs, & Chi, 2016). Hence, another reason for finding more Microsoft Academic citations to books is the current BKCI problem with linking existing citations from WoS to BKCI-indexed books.

Table 1. Examples of books without BKCI citations but with WoS citations

| Example 1 | 44. Title: [not available]. By: Hurd, Elizabeth Shakman. Beyond Religious Freedom: The New Global Politics of Religion Published: 2015 Publisher: Princeton University Press, Princeton, NJ  **Times Cited: 57** *(from Web of Science Core Collection)* |
|---|---|
| Example 2 | Title: [not available]. By: Ansdell, G. **How Music Helps in Music Therapy and Everyday Life**. Published: 2014. Publisher: Ashgate Publishing Limited., Surrey. |
| Example 3 | Introduction: A primer on information and influence in animal communication. By: Stegmann, Ulrich E. Edited by: Stegmann, UE. **ANIMAL COMMUNICATION THEORY: INFORMATION AND INFLUENCE.** Pages: 1-39 Published: 2013. |

### 6.1.3. Citations to different book editions

Microsoft Academic includes citations to different editions of a book (if any). It systematically matches authors and titles of cited books, irrespective of the edition. It may use a similar matching strategy to that for articles (Sinha et al., 2015) which helps to integrate citations to preprint and published versions of the same paper (Kousha & Thelwall, 2018). For example, the second edition of book "*Biotechnology of Lactic Acid Bacteria: Novel Applications*" published in 2015 by Fernanda Mozzi had 86 Microsoft Academic citations. However, manual checks showed that 30 citations to this book were to the earlier (first) edition published in 2010. In contrast, BKCI is edition-sensitive, assigning citations to different editions of a book separately (Gorraiz, Purnell, & Glänzel, 2013; Glänzel, Thijs, & Chi, 2016), reducing the citation counts for each individual edition. Integrating citations to different editions of books could be useful when the overall impact of a book needs to be assessed.

### 6.2. Limitations

Overall, since only 60% of the BKCI books checked were found by Microsoft Academic, its coverage of books is not comprehensive and may not be extensive. In the current study, for books with multiple editions, only the Microsoft Academic citations to the edition in BKCI were counted to give comparable results. However, Microsoft Academic's coverage of BKCI Books would be higher if different book editions were included. Moreover, Microsoft Academic sometimes collects citations from different editions of a book. This could be problematic when the different versions are substantially different, although there is not a recommended solution to this problem (Gorraiz, Purnell, & Glänzel, 2013; Glänzel, Thijs, & Chi, 2016).

The results to the second and third research questions may be influenced by the absence of records for many of the BKCI books. Presumably, Microsoft Academic only indexes books that it finds a publisher record for, but it seems possible that these would be more (or less) cited than the books that it has no record for.

Whilst for retrieving Microsoft Academic records for journal articles, false matches can be filtered out using article titles, authors, journal names, DOIs and publication years (Thelwall, 2018b), books lack DOIs and journal years, reducing filtering effectiveness. In addition, ISBNs and publisher names are not included in most Microsoft Academic records and so cannot be used. Hence, in some cases the filtering strategy may retain incorrect matches, especially for books with general or short titles (e.g., *Introduction to Psychology*) and common author names (e.g., Smith or Zhang). The title, authors/editors and publication year filtering may also remove some correct matches (Thelwall, 2018b). This could occur for books with incorrect publication years in BKCI or Microsoft Academic (Harzing & Alakangas, 2017b; Hug, Ochsner, & Brändle, 2017). For instance, although Microsoft Academic shows the correct publication year (2015) of the book "*Biomass Sugars for Non-Fuel Applications*" by Dmitry Murzin, BKCI incorrectly reports 2016. Similarly, the book "*Phantom Limbs: On Musical Bodies*" by Peter Szendy has the correct publication year in BKCI (2016) but is incorrect in Microsoft Academic (2015). There was no practical method to include correct matches for these cases and to remove results from different book editions for a fairer comparison between BKCI and Microsoft Academic.

Finally, the coverage of Microsoft Academic of non-English books and international publishers is not known because the BKCI data used is dominated by English books from the UK and the USA (Torres-Salinas et al., 2014). Thus, it is not clear whether Microsoft Academic is useful for non-English books or books published in other countries.

### 7. Conclusions

Although Microsoft Academic does not claim to be a citation index for books, it incorporates records for at least 60% of BKCI books from 2013-2016 and has extracted some citations from books in addition to citations from journal articles, preprints and other document types. It has better coverage of older books and scientific books. Since BKCI is itself a small subset of academic books from prestigious publishers, in answer to the first research question, the current version of Microsoft Academic seems to have very incomplete coverage. Its coverage is lower in the book-based areas Art (48%), History and Archaeology (50%), Educational Sciences (50%), and Philosophy, Ethics and Religion (55%) and higher in Computer Science (79%), Chemical Sciences (77%) and Physics and Mathematics (72%).

In answer to the second research question, Microsoft Academic captured 1.5 to 3.6 times more citations than BCKI in 9 out of 17 fields during 2013-2016, whereas BKCI citation counts were not statistically higher than Microsoft Academic only for *Sociology and History and Archaeology* in 2013 and 2014 and for *Computer Science* in 2013. This suggests that Microsoft Academic could be a useful alternative source of citations for books that it indexes or when BKCI is not available for evaluators, research committees or funders. It may be useful for research evaluations when comprehensive coverage is not needed but a sample could be used instead. The citation advantage of Microsoft Academic over BKCI was partly due to BKCI being unable to match some existing WoS citations to BKCI books. It was partly also due to Microsoft Academic finding many non-WoS citations to books in *History and Archaeology* (76%), *Art* (73%), *Languages and Literature* (71%) and *Law* (70%) and *Sociology* (70%). Microsoft Academic had a lower citation advantage over Google Books. Average Microsoft Academic citation counts were higher than Google Books in six out of 17 fields, whereas Google Books had a citation advantage over Microsoft Academic in four subject areas, suggesting that the two databases have partly complementary coverage.

In answer to the third research question, Microsoft Academic and BKCI citation counts for BKCI books correlate positively and statistically significantly across all subject areas and years (except for *Philosophy, Ethics and Religion* and *Other Social Sciences* in 2016). The correlations between the Microsoft Academic and BKCI citations are much higher in sciences than other fields, perhaps because in science subject areas many citations came from articles indexed by both Microsoft Academic and WoS. However, there are lower (and sometimes statistically insignificant) correlations between BKCI and Google Books citations across most fields than between Microsoft Academic and Google Books, suggesting that Microsoft Academic partly reflects the book-based impact that can be found in Google Books. This is possible because it extracts citations *from* some books.

Microsoft Academic's ability to extract citations from sources not indexed by BKCI seems to be an advantage for citation analysis of academic books, although it seems to have too few records for academic books to be useful for comprehensive research evaluations. Microsoft Academic was mainly designed for searching scholarly journal articles and conference papers and there seems to be no policy for indexing academic books (https://academic.microsoft.com/#/faq). Nevertheless, Microsoft Academic's coverage of scholarly literature increased from 127 million in February 2016 (Herrmannova & Knoth, 2016) to 168 million in early 2017 (Hug & Brändle, 2017) and indexed about 175 million publications in June 2018 (https//academic.microsoft.com/). Hence, its coverage of academic books might also increase over time (Harzing & Alakangas, 2017b), presumably through finding recently published works, but it may expand more extensively if open access monograph initiatives (www.hefce.ac.uk/pubs/rereports/year/2015/monographs) are successful.